\documentclass[%
aip,
amsmath,amssymb,
 reprint,%
]{revtex4-1}

\usepackage{graphicx}
\usepackage{dcolumn}
\usepackage{bm}

\usepackage[utf8]{inputenc}
\usepackage[T1]{fontenc}
\usepackage{mathptmx}
\usepackage{etoolbox}

\makeatletter
\def\@email#1#2{%
 \endgroup
 \patchcmd{\titleblock@produce}
  {\frontmatter@RRAPformat}
  {\frontmatter@RRAPformat{\produce@RRAP{*#1\href{mailto:#2}{#2}}}\frontmatter@RRAPformat}
  {}{}
}%
\makeatother
\begin{document}

\preprint{AIP/123-QED}

%
%
%
%

\title[Design of a multifunctional Doppler backscattering diagnostic for the Pegasus-III Experiment]{Design of a multifunctional Doppler backscattering diagnostic for the Pegasus-III Experiment}

\author{E. Wikarta}
\email{Eduard\_Wikarta@a-star.edu.sg}
\affiliation{Future Energy Acceleration and Translation (FEAT) Centre, Agency for Science, Technology and Research (A*STAR), Singapore 138632, Singapore}

\author{U. Kumar}
\affiliation{Future Energy Acceleration and Translation (FEAT) Centre, Agency for Science, Technology and Research (A*STAR), Singapore 138632, Singapore}

\author{V.H. Hall-Chen}
\affiliation{Future Energy Acceleration and Translation (FEAT) Centre, Agency for Science, Technology and Research (A*STAR), Singapore 138632, Singapore}
\affiliation{School of Physical and Mathematical Sciences, Nanyang Technological University}

\author{S.J. Diem}
\affiliation{Department of Nuclear Engineering and Engineering Physics, University of Wisconsin-Madison, WI, USA}

\author{X. Li}
\affiliation{Future Energy Acceleration and Translation (FEAT) Centre, Agency for Science, Technology and Research (A*STAR), Singapore 138632, Singapore}

\author{K.T.E. Chua}
\affiliation{Future Energy Acceleration and Translation (FEAT) Centre, Agency for Science, Technology and Research (A*STAR), Singapore 138632, Singapore}

\author{T.S.P. See}
\affiliation{Future Energy Acceleration and Translation (FEAT) Centre, Agency for Science, Technology and Research (A*STAR), Singapore 138632, Singapore}

\author{A.C. Sontag}
\affiliation{Department of Nuclear Engineering and Engineering Physics, University of Wisconsin-Madison, WI, USA}

\author{Z. Wilderspin}
\affiliation{Department of Nuclear Engineering and Engineering Physics, University of Wisconsin-Madison, WI, USA}

\date{\today} 

\begin{abstract}
\textbf{Abstract} The Doppler backscattering (DBS) diagnostic measures flows and electron density fluctuations. Recent work indicates that DBS can also infer the magnetic pitch angle (Yeoh et al., NF 2026). We present the preliminary design of a DBS for the Pegasus-III Experiment. This DBS will serve two objectives. First, it will advance diagnostic science, by supporting understanding of the DBS instrumentation functions, using DBS to constrain the magnetic equilibrium, and data-driven inference of plasma properties from DBS signals. Secondly, it will support Pegasus-III’s research directions, such as solenoid-free plasma initiation and O-X-B mode conversion for heating and current drive, since density fluctuations affect mode conversion efficiency and pitch angle measurements can be used to locate the mode conversion window. This ex-vacuum DBS system uses a single channel, tuneable Ka-band source, a corrugated horn antenna, and a homodyne I/Q receiver with baseband digitization. For greater flexibility, which is especially important for pitch angle measurements, the quasioptical elements include a rotatable spinner for O- and X-mode selection and a mirror with 2D steering. Using the \textit{Scotty} beam-tracing code, for a range of poloidal launch angles $8^\circ$ to $18^\circ$ and a corresponding toroidal launch angle between $0^\circ$ to $3^\circ$ for maximal backscattered DBS power, we find that the DBS system is capable of measuring ion-scale density fluctuations $1\leq k_{\perp,c} \leq8 \text{ cm}^{-1}$ at a range of normalized radial coordinates from the outer core ($\rho \sim 0.65$) to just beyond the last-closed flux surface (LCFS), where $\rho=0$ corresponds to the magnetic axis and $\rho=1$ the LCFS. The system is also designed with additional toroidal steering capability, $-4^\circ$ to $8^\circ$, to resolve the toroidal response needed for magnetic pitch angle measurements.
\end{abstract}

\maketitle


%
%
\section{Introduction}
Plasma initiation in a tokamak typically relies on the central solenoid inductively driving a plasma current, which is not ideal for fusion power plants because this solenoid occupies valuable central-column space, making neutron shielding more challenging and reducing the real-estate available for tritium-breeding. This constraint is especially severe in spherical tokamaks, where the central column is even narrower than in conventional tokamaks. Solenoid-free initiation is therefore crucial for spherical-tokamak power plants and desireable for conventional-tokamak power plants \cite{Schaefer:LHI:2026}. Such initiation is the primary focus of Pegasus-III, a low-aspect-ratio spherical tokamak, which features non-inductive heating and current drive (H\&CD) techniques, such as local helicity injection (LHI) and electron Bernstein waves (EBW) via O-X-B mode conversion \cite{Sontag:Pegasus-III:2022, Weberski:LHI:2025, Goetz:EBW:2026}. Among the measurements required to understand non-inductively driven plasmas are density fluctuations, plasma flows, and magnetic-field geometry, for two reasons:
\begin{itemize}
    \item During LHI, biased edge injectors drive current along open magnetic field lines. As these current streams reconnect and relax, they perturb the edge magnetic topology and fluctuation environment \cite{Richner:LHI:2022, Weberski:LHI:2025, Schaefer:LHI:2026}.
    \item During EBW H\&CD, the O-X-B mode conversion strongly depends on the density profile, density fluctuations, and magnetic-field geometry near the mode-conversion layer \cite{Goetz:EBW:2026}.
\end{itemize}

Pegasus-III already has several diagnostics relevant to these objectives, including Thomson scattering for electron density and temperature profiles \cite{Schlossberg:Thomson:2016}, magnetic diagnostics for equilibrium reconstruction and local magnetic-fluctuation measurements\cite{Schaefer:magnetics:2026}, and synthetic-aperture microwave imaging (SAMI) \cite{Freethy:SAMI:2013, Thomas:SAMI:2016, Peery:EBW:2024}. Through SAMI's broad emission pattern and two-dimensional imaging capability, it can, in principle, measure density fluctuations, flows, and the magnetic pitch angle. However, interpreting it generally requires computationally intensive full-wave simulations \cite{Thomas:SAMI:2016}. Hence, there is a need for a complementary diagnostic that provides routinely interpretable measurements of the same properties, with a measurement range that overlaps SAMI in the edge while extending deeper into the core.

Doppler backscattering (DBS) is well suited to this role. DBS is a microwave diagnostic traditionally used to measure turbulent electron density fluctuations and flows \cite{Hennequin:DBS:2006, Happel:DBS:2009, Pratt:DBS:2022, Rhodes:DBS:2022, Shi:DBS:2023, Chowdhury:DBS:2023, Macwan:DBS:2024, Shi:DBS:2025, Tong:DBS:2025, Tokuzawa:DBS:2025, Liang:DBS:2026}. A microwave probe beam is launched into the plasma and scatters from density fluctuations. The backscattered power is measured, which is typically dominated by scattering from the cutoff location. The received signal's Doppler shift is related to the flow velocity, while the backscattered amplitude is proportional to the density fluctuation amplitude. The cutoff location is predominantly determined by the probe beam's frequency, while the launch angle controls the measured fluctuation wavenumber through the Bragg condition\cite{Pratt:DBS:2024}. 

Additionally, DBS can also be used to infer the magnetic pitch angle. The backscattered power is maximised when the probe beam's wavevector is perpendicular to the magnetic field at the scattering location. By measuring the dependence of backscattered power on toroidal launch angle, referred here to as the toroidal response, one can infer the local magnetic pitch angle, which was demonstrated at the DIII-D conventional tokamak \cite{Yeoh:DBS:2025}. Finally, DBS can yield additional plasma properties through data-drive techniques \cite{Teo:ELM:2024}. Although such techniques are not considered in this paper, they provide a possible extension of this diagnostic's capability in the future.

In this paper, we present the preliminary design of a two-dimensionally steerable DBS diagnostic for Pegasus-III. The system uses a single-channel tuneable Ka-band source, a corrugated horn antenna, a spinner for O- and X-mode selection, and a two-dimensionally steerable quasioptical launcher. We show that this system can operate as a conventional DBS diagnostic, accessing useful scattering locations and fluctuation wavenumbers in Pegasus-III while satisfying engineering constraints. We also show that the same hardware can provide a test-bed for pitch angle DBS in a spherical tokamak. The rest of this paper is structured as follows. Section~\ref{sec:Pegasus-III_equilibria_and_DBS_requirements} describes the Pegasus-III plasma scenarios and diagnostic requirements. Section~\ref{sec:beam-tracing_assessment_of_DBS_performance} presents beam-tracing calculations of cutoff locations, measured wavenumbers, and mismatch attenuation. Section~\ref{sec:quasioptical_and_hardware_design} describes the preliminary quasioptics, steering design, and microwave hardware.

\section{Pegasus-III equilibria and launch geometry} \label{sec:Pegasus-III_equilibria_and_DBS_requirements}
%
%
We selected two representative L-mode Pegasus-III plasma scenarios to assess the DBS design requirements and measurement capabilities.

We considered a DBS diagnostic where the probe beam is launched from the large midplane port, with the steering mirror located just outside the port window (see Figure \ref{fig:port_window}). The large midplane port was chosen because it provides sufficient line-of-sight access to the plasma edge and core while while accommodating the required range of poloidal and toroidal steering, given in the next section. Depending on the magnetic field coil set, the magnetic axis can either be at or below the geometric midplane. Hence, the DBS is designed to launch slightly above the midplane with predominantly downward poloidal steering, denoted with a positive sign in \textit{Scotty} convention. Similarly, a positive toroidal launch angle corresponds to steering leftwards.

\begin{figure}
    \centering
    \includegraphics[width=0.47\textwidth]{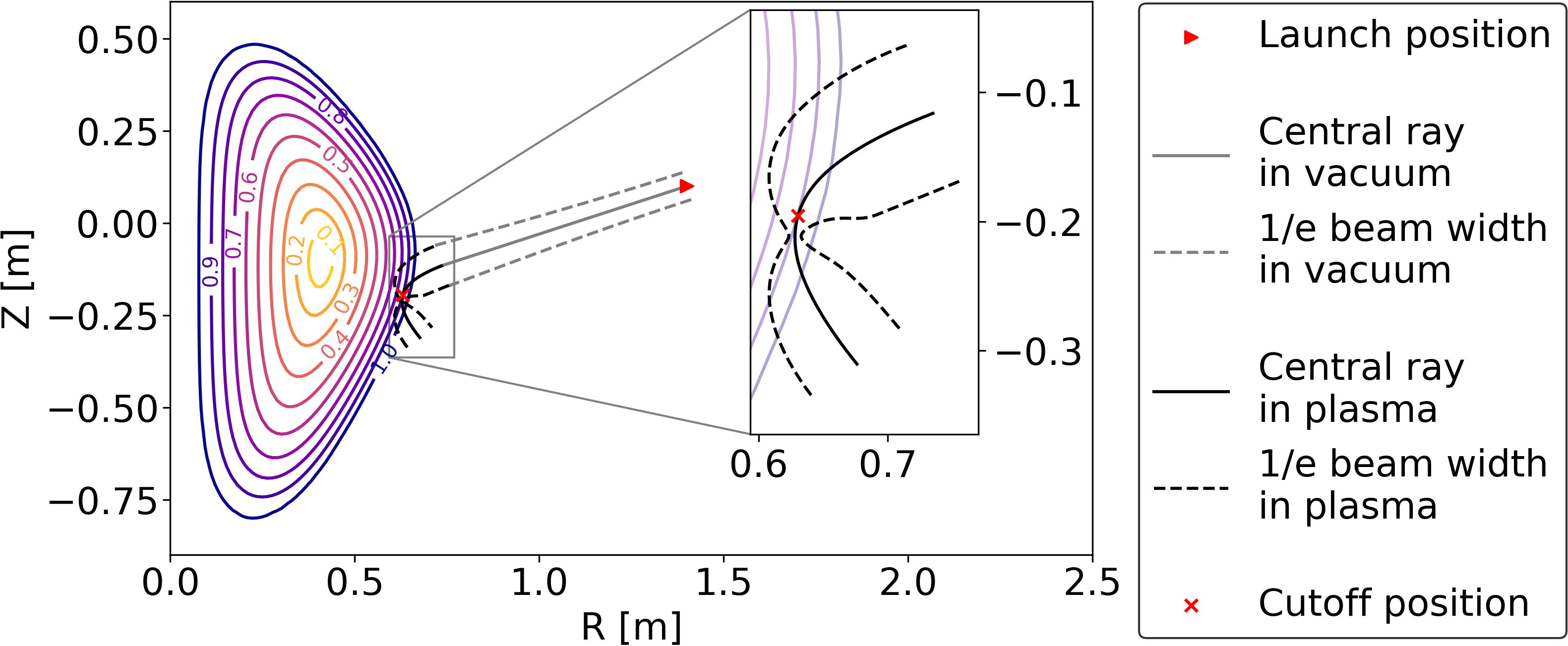}
    \caption{\textit{Scotty} beam-tracing simulation of a $f=40$ GHz probe beam propagating through the lower-density Pegasus-III plasma, shot P9630, during flat top. Contours show the flux surfaces. The poloidal and toroidal launch angles are $(\varphi_p, \ \varphi_t) = (18.0^\circ, 1.8^\circ)$, which achieve low mismatch at the cutoff, $|\theta_{m,c}|\sim0^\circ$. The red triangle shows the beam's launch position at the steering mirror, the solid black line shows the trajectory of the central ray, and the dashed lines show the $1/e$ beam width of the electric field. The beam width narrows near the cutoff, increasing the electric field and contributing to the dominance of the cutoff location in the received signal. As Pegasus-III plasmas are typically shifted downward relative to the midplane port, we focused on downward poloidal steering.}
    \label{fig:beam_tracing_for_P9630_equilibrium}
\end{figure}

Beam trajectories were calculated using the \textit{Scotty} beam-tracing code \cite{Hall-Chen:beam:2022} for two experimentally realised L-mode Pegasus-III plasmas during flat top, see Figure \ref{fig:beam_tracing_for_P9630_equilibrium}. In particular, we used two shots:
\begin{itemize}
    \item Shot P9630, a lower-density plasma, $n_{e,0} = 3.9 \times 10^{19} \textrm{m}^{-3}$, where $n_{e,0}$ is the core electron density.
    \item Shot P7285, a higher-density plasma, $n_{e,0} = 2.6 \times 10^{19} \textrm{m}^{-3}$.
\end{itemize}
These plasmas span a range of Pegasus-III operating densities relevant to LHI and EBW studies. The magnetic equilibria and density profiles were obtained from magnetics-constrained Grad-Shafranov reconstructions\cite{Weberski:equilibrium:2024} and Thomson scattering measurements \cite{Schlossberg:Thomson:2016}. For input into \textit{Scotty}, the density profiles from Thomson scattering measurements were fitted with polynomial functions,
\begin{align*}
    n_{e,high}(\rho)  &= -2.885\rho + 3.918, \\
    n_{e,low}(\rho) &= -1.303\rho^2 + 2.589,
\end{align*}
where $0\leq\rho\leq1$ is the normalised radial coordinate, with $\rho=0$ and $\rho=1$ corresponding to the magnetic axis and last-closed flux surface, respectively, and the electron density is in units of $\times10^{19}\text{ m}^{-3}$.

\section{Beam-tracing assessment of DBS performance} \label{sec:beam-tracing_assessment_of_DBS_performance}
%
%
We proceed to determine the range of frequencies, polarisations, as well as toroidal and poloidal steering required for the Pegasus DBS to function as both a conventional turbulence diagnostic and a pitch angle diagnostic.

\subsection{Measurement locations and wavenumbers} \label{subsec:measurement_locations_and_wavenumbers}
In DBS, the received signal is typically dominated by scattering from density fluctuations near the cutoff location. We therefore use \textit{Scotty} beam-tracing simulations to determine the cutoff locations for a range of launch frequencies, launch angles, and both O- and X-mode polarisations. The fluctuation wavenumber measured by DBS is given by the Bragg condition and is approximately twice the probe beam wavenumber at the scattering location. Using \textit{Scotty}, we calculate the probe beam wavevector at the cutoff and thus the measured fluctuation wavenumber.

The measured fluctuation wavenumber depends strongly on the poloidal launch angle, which is bounded at both shallow and steep ends. At shallow launch angles, around $8^\circ$, the inward- and outward-propagating beams overlap significantly, adversely affecting the applicability of beam tracing \cite{Ruiz:beam:2025}. At even shallower launch angles, the probe beam becomes nearly normal to the flux surfaces, and the measurement approaches the conventional reflectometry regime rather than the DBS regime. In this limit, density fluctuations may still be measurable, but Doppler-shifted measurements become more difficult. On the other hand, steeper launch angles provide access to larger fluctuation wavenumbers and broaden the measurable turbulence range. However, the weaker refraction at these larger launch angles results in poorer beam focusing near the cutoff, and thus a lower spatial resolution due to sensitivity to scattering away from the cutoff location \cite{Ruiz:beam:2025}. Larger fluctuation wavenumbers are also more difficult to measure because of sensitivity to mismatch attenuation \cite{Hillesheim:DBS:2015, Hall-Chen:validation:2022} and lower fluctuation amplitudes \cite{Barnes:spectrum:2011}. Based on these considerations, we selected a poloidal launch angle range of $8^\circ\leq\varphi_p\leq18^\circ$.

\begin{figure}
    \includegraphics[width=0.47\textwidth]{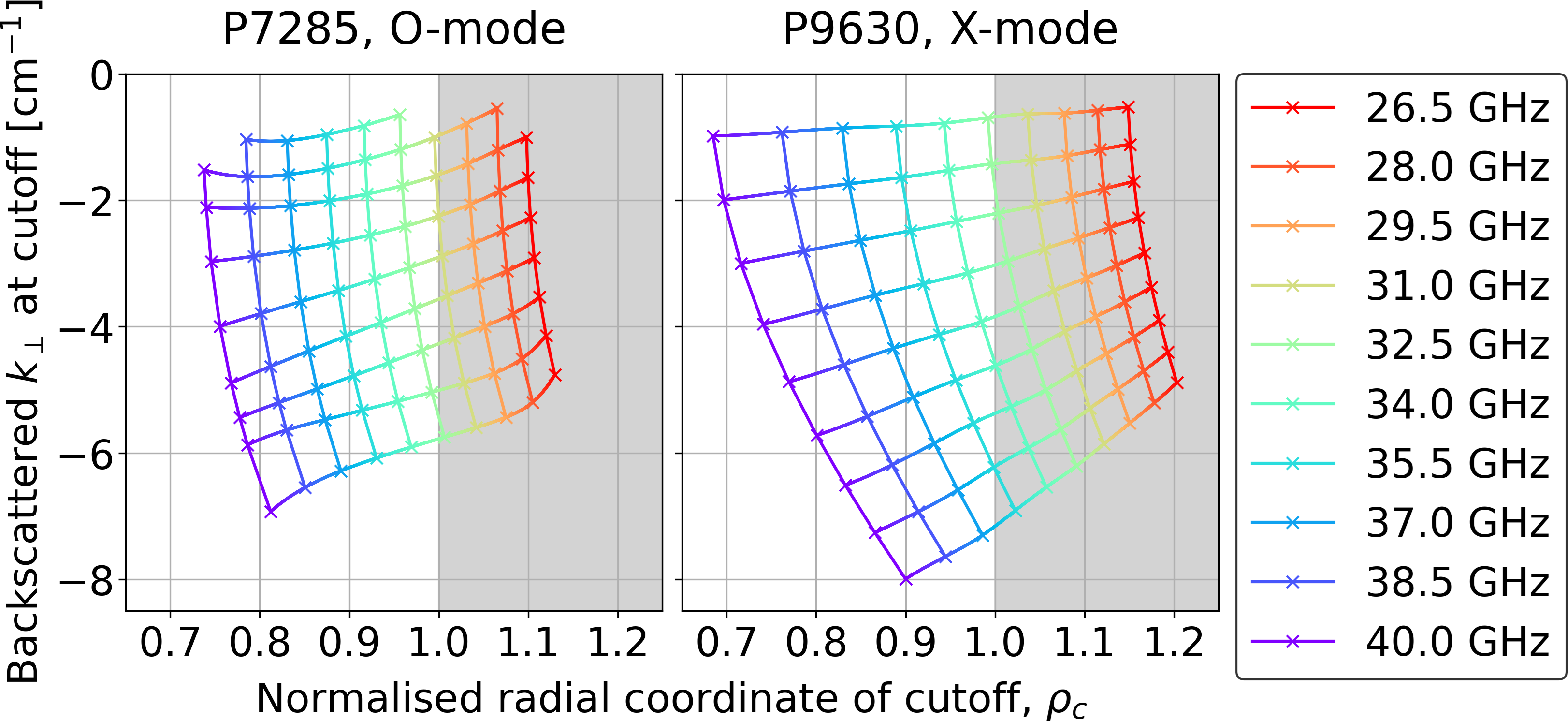}
    \caption{DBS measurement locations and measured turbulence wavenumbers obtained from \textit{Scotty} beam-tracing scans over probe frequency and poloidal launch angle, plotted at intervals of $\varphi_p = 1^\circ$, with the same steepest angle for both subplots, $\varphi_p = 18^\circ$. Results are shown for the high-density equilibrium with O-mode polarisation (left) and the low-density equilibrium with X-mode polarisation (right). Here, $k_{\perp,c}$ is the measured turbulence wavenumber at the cutoff and $\rho_c$ is the normalised radial coordinate of the cutoff location. A Ka-band system with poloidal steering, $8^\circ\leq\varphi_p\leq18^\circ$, can access a range of wavenumbers, $1 \lesssim k_{\perp,c}\lesssim 8\text{ cm}^{-1}$, from $\rho = 0.65$ to the last-closed flux surface.}
    \label{fig:kperp_vs_rho_plot}
\end{figure}

With \textit{Scotty} parameter scans over launch frequency, poloidal launch angle, equilibrium, and probe beam polarisation, we find that a Ka-band ($26.5$--$40$ GHz) DBS system and $10^\circ$ of poloidal steering, $8^\circ\leq\varphi_p\leq18^\circ$, can access radial locations from the outer core to the edge, $0.65\lesssim\rho\leq1$, over a range of turbulence wavenumbers, $1\lesssim k_{\perp,c}\lesssim 8\text{ cm}^{-1}$, see Figure \ref{fig:kperp_vs_rho_plot}. This corresponds to $0.5 \lesssim k_\perp\rho_s \lesssim 3.5$, covering ion-scale turbulence. Here $\rho_s$ is the ion sound gyroradius. To enable operation across a range of plasma scenarios, the system is designed to switch between O- and X-mode probe beam polarisations. The selected Ka-band range also extends to somewhat lower frequencies than required for the flat-top plasmas considered here, retaining flexibility for future measurements during lower-density startup and ramp-up phases, although such plasmas are not explicitly studied in this paper.

\subsection{Toroidal steering for turbulence and pitch angle measurements}
This DBS system requires toroidal steering for pitch angle matching for turbulence measurements, as the pitch angle is large and varies spatially in Pegasus-III. Toroidal steering is especially important for measurements of larger turbulence wavenumbers \cite{Hillesheim:DBS:2015, Damba:DBS:2022, Pratt:DBS:2024}. As most turbulent density fluctuations are highly anisotropic with small parallel wavenumbers, they exist primarily in the plane perpendicular to the magnetic field. Hence, the wavevector of the probe beam must be aligned with this plane for optimal backscattering. Misalignment, or mismatch, decreases the backscattered signal, even when the density fluctuation amplitude is kept unchanged. Moreover, toroidal steering is crucial for pitch angle measurements \cite{Yeoh:DBS:2025}, as this technique depends on the toroidal response to calculate the pitch angle. The rest of this subsection is dedicated to determining the range of toroidal steering required.

\begin{figure}
    \includegraphics[width=0.47\textwidth]{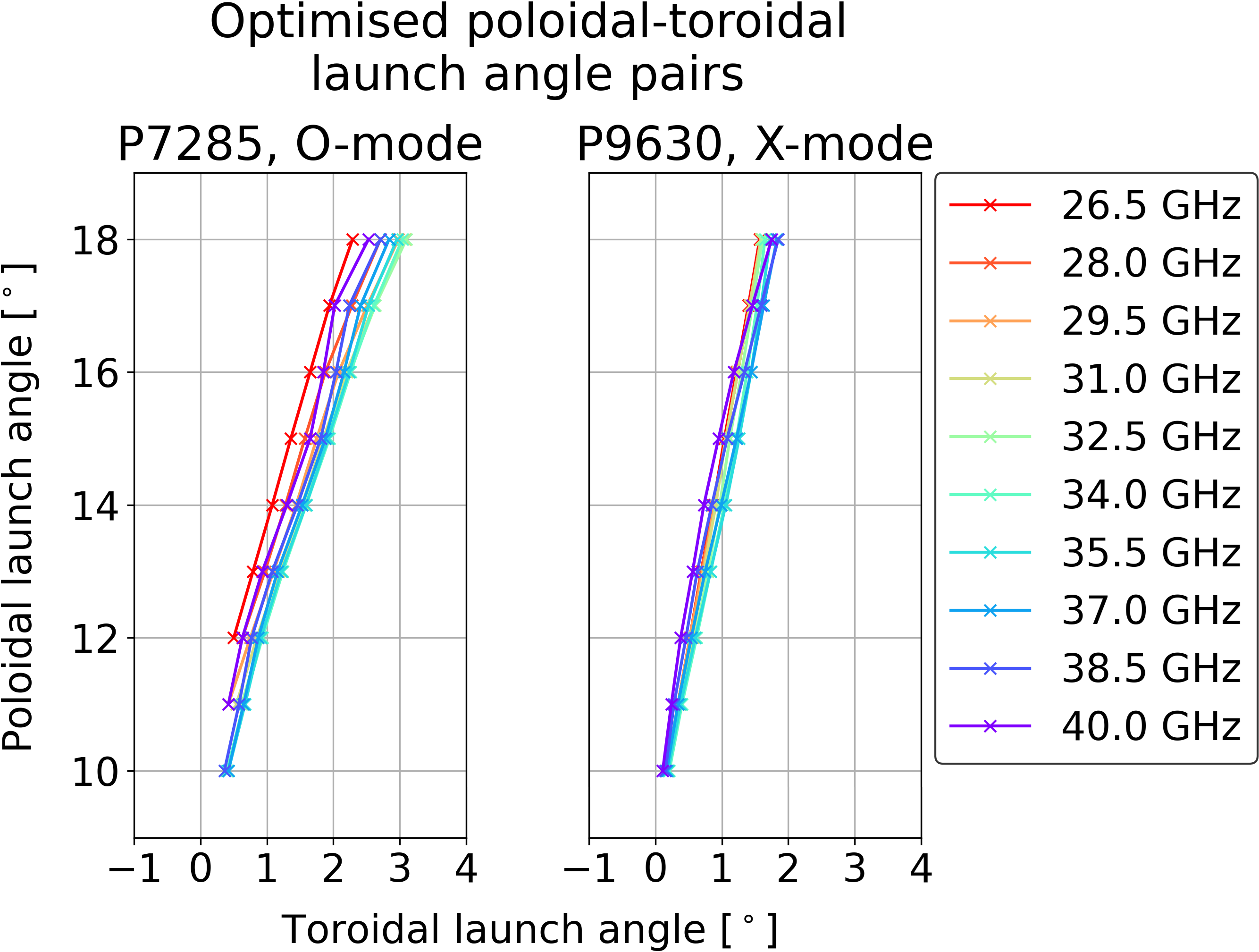}
    \caption{Contour lines of pitch angle matching at cutoff for various frequencies for the same equilibria in Figure~\ref{fig:kperp_vs_rho_plot}. Here, each contour line denotes the poloidal-toroidal launch angle pairs $(\varphi_p, \ \varphi_t)$ required for matching with the field line at the cutoff, where in this work we assume most of the received signal comes from. These angles thus yield an estimate of the range of toroidal steering needed, which is used later to give an estimate of the port size required.}
    \label{fig:pol_tor_plots}
\end{figure}

Given the range of poloidal launch angles in Section \ref{subsec:measurement_locations_and_wavenumbers}, we calculated the corresponding toroidal launch angles required for pitch angle matching at the cutoff (see Figure~\ref{fig:pol_tor_plots}). We found that a toroidal range of $0^\circ$ to $3^\circ$ is required to achieve pitch angle matching for all frequencies.

\begin{figure}
    \includegraphics[width=0.45\textwidth]{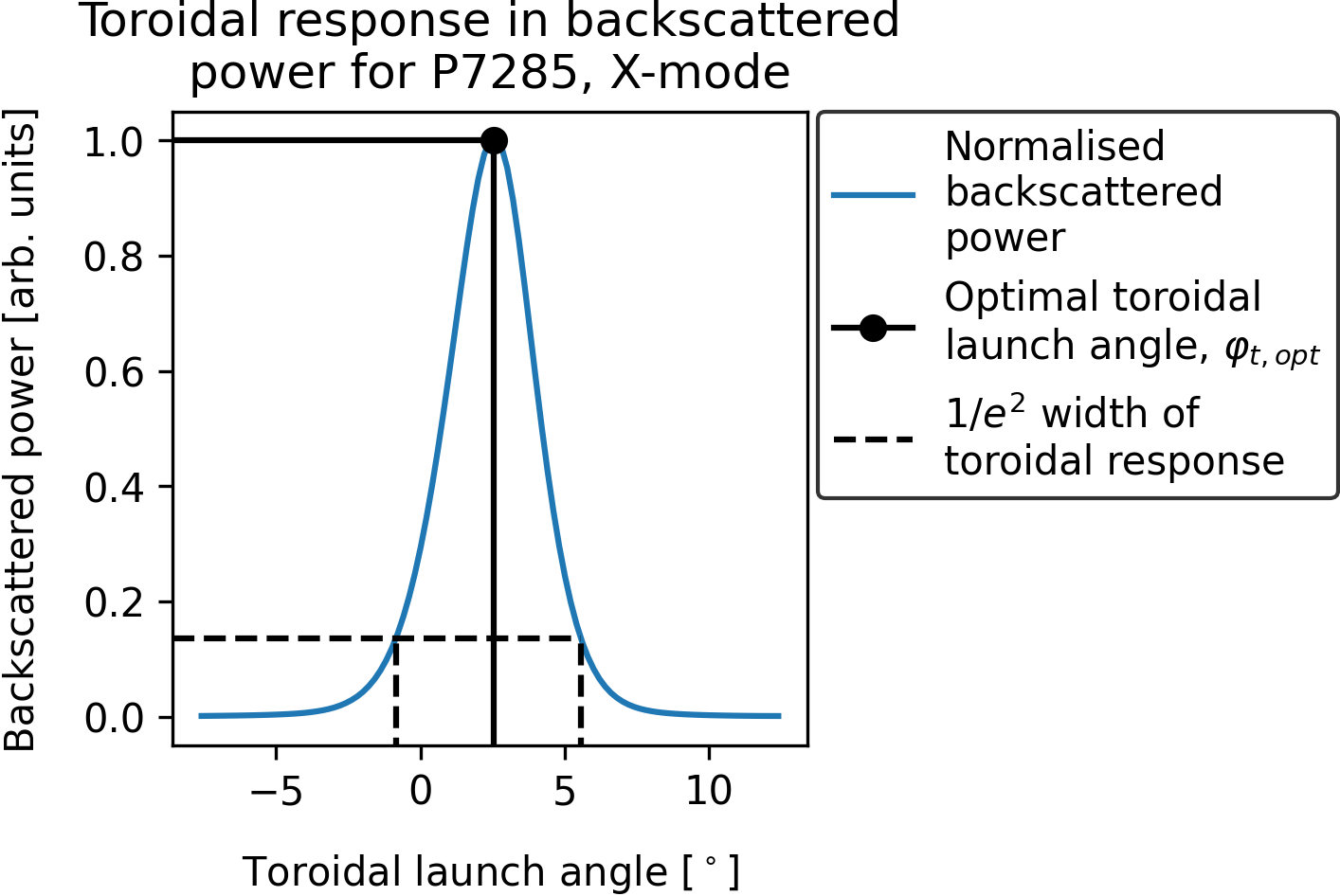}
    \caption{Taking the left-most and right-most points (i.e. the minimum and maximum toroidal launch angles) of Figure~\ref{fig:pol_tor_plots} yields the greatest extent of the toroidal steering required for the DBS system. For these angles, we find their mismatch tolerance \cite{Hall-Chen:beam:2022} by sweeping a range of toroidal angles with a fixed poloidal angle and then finding the $1/e^2$ width of the toroidal response. Note that only one of the toroidal response graphs is depicted here.}
    \label{fig:mismatch_tolerance_plots}
\end{figure}

Compared to conventional turbulence DBS, pitch angle DBS requires a larger range of toroidal steering. First, the peak of the toroidal response must be resolved, so the diagnostic must measure the fall-off in backscattered power away from the launch angle that satisfies pitch angle matching, while turbulence DBS only requires that matching is achieved. We use \textit{Scotty} to predict the toroidal response and obtain the toroidal range needed to see a drop in the received power by a factor of $1/e^2$, see Figure \ref{fig:mismatch_tolerance_plots}. This consideration results in a large range of toroidal launch angles from $-2^\circ$ to $6^\circ$.

Furthermore, the actual magnetic pitch angle may differ from the magnetics-constrained equilibrium used in \textit{Scotty}. We therefore expand the toroidal steering range as a contingency measure. Consequently, our DBS design has a toroidal steering range of $-4^\circ$ to $8^\circ$.

\begin{figure}
    \includegraphics[width=0.47\textwidth]{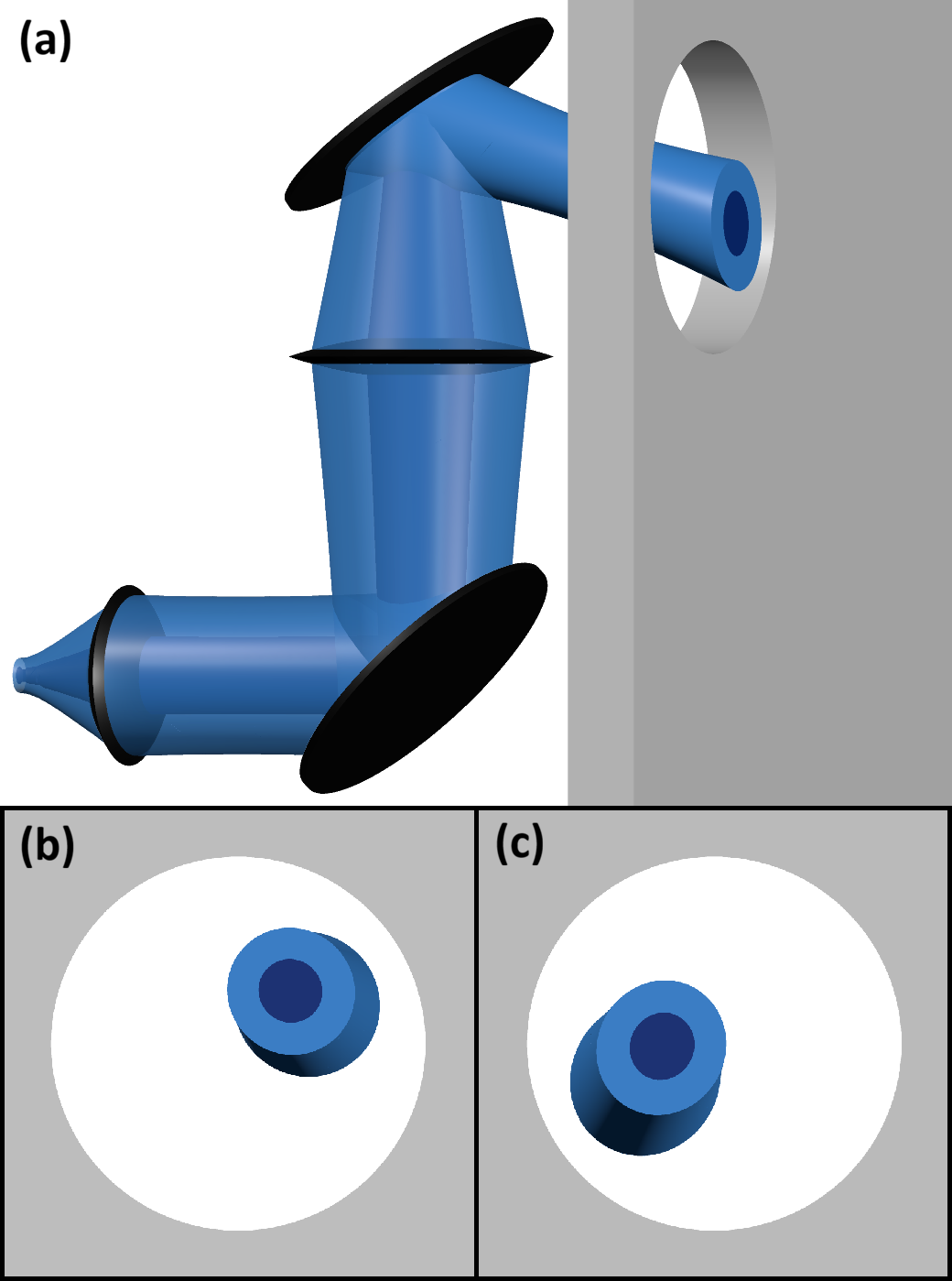}
    \caption{With the poloidal-toroidal steering found, we find an estimate for the port window based on the requirement that the 2-times-beam radius fits completely within the port \cite{Goldsmith:Quasioptics:1998}. Given the quasioptical set-up in Section \ref{sec:quasioptical_and_hardware_design}, we calculate the trajectory of the beam from antenna to port: (a) in an isometric perspective; and the inward (to the plasma) perspective for (b) $(\varphi_p, \varphi_t) = (8^\circ, 10^\circ)$; and (c) $(\varphi_p, \varphi_t) = (18^\circ, -10^\circ)$. We see that the port window is able to comfortably accommodate the required steering.}
    \label{fig:port_window}
\end{figure}

Finally, we verify that the required range of poloidal and toroidal launch angles can indeed fit into the port window. We calculate the beam width at the window from the quasioptical system, detailed below, accounting for the oblique incidence, and find that the port window is indeed sufficiently large, see Figure \ref{fig:port_window}.

\section{Quasioptical and hardware design} \label{sec:quasioptical_and_hardware_design}
The beam-tracing results above require a DBS system with Ka-band operation, switchable O- and X-mode polarisation, and two-dimensional steering over a wider toroidal range than is necessary for conventional DBS. The hardware must also produce a well-characterised Gaussian beam, since the launch position, beam width, and beam curvature are used directly in the \textit{Scotty} calculations. We therefore designed an ex-vacuum quasioptical system that provides the required steering range while allowing the launch angles and beam properties to be calibrated before installation.

\subsection{Quasioptical design}
The quasioptical design is constrained by three main requirements. First, the system must produce a well-approximated Gaussian beam for beam-tracing interpretation, with limited clipping by optical elements and the port aperture. Second, the beam incident on the plasma must remain sufficiently focused to provide useful spatial localisation near the cutoff. Third, the final steering element must be close enough to the port window to provide the required poloidal and toroidal steering range.

These requirements lead to a lens, fixed-mirror, lens, steering-mirror layout. The first lens partially collimates the beam from the corrugated horn, reducing beam expansion through the intermediate optical path. A fixed mirror folds the beam path, allowing the second lens and steering mirror to fit within the available ex-vacuum space. The second lens focuses the beam toward the plasma, while the final steering mirror controls the poloidal and toroidal launch angles. The resulting beam parameters are used as inputs to the \textit{Scotty} calculations, and the principal optical dimensions are summarised in Table~\ref{tab:horn_and_lens_specifications}.
\begin{table}
    \begin{tabular}{|l|l|l|}
    \hline
    \textbf{Component} & \textbf{Details} & \textbf{Linear distance} \\
    \hline
    Horn & Linear-polarised, corrugated & -- \\
    Lens & UHMWPE, $f=0.15$ m & 0.15 m \\
    Fixed mirror & Flat, radius 0.20 m & 0.40 m \\
    Lens & UHMWPE, $f=0.45$ m & 0.40 m \\
    Steering mirror & Flat, radius 0.20 m & 0.30 m \\
    \hline
    \end{tabular}
    \caption{Horn and lens specifications used for the quasioptical calculations shown in Fig.~\ref{fig:port_window} and beam-tracing simulations. The refractive index of UHMWPE \cite{Xie:JTEXT-ECE:2020}, $n=1.575$, is assumed constant.}
    \label{tab:horn_and_lens_specifications}
\end{table}

We place the steering mirror close to the port window to preserve angular access through the port aperture. This is particularly important for Pegasus-III because toroidal steering is required both to minimise mismatch attenuation during turbulence measurements and to resolve the toroidal response for pitch angle DBS. The ex-vacuum layout also allows the beam direction and beam profile to be measured before installation, providing a practical route to calibrating the launch angles and Gaussian beam parameters.

\subsection{Microwave source and receiver chain}
\begin{figure}
    \includegraphics[width=0.47\textwidth]{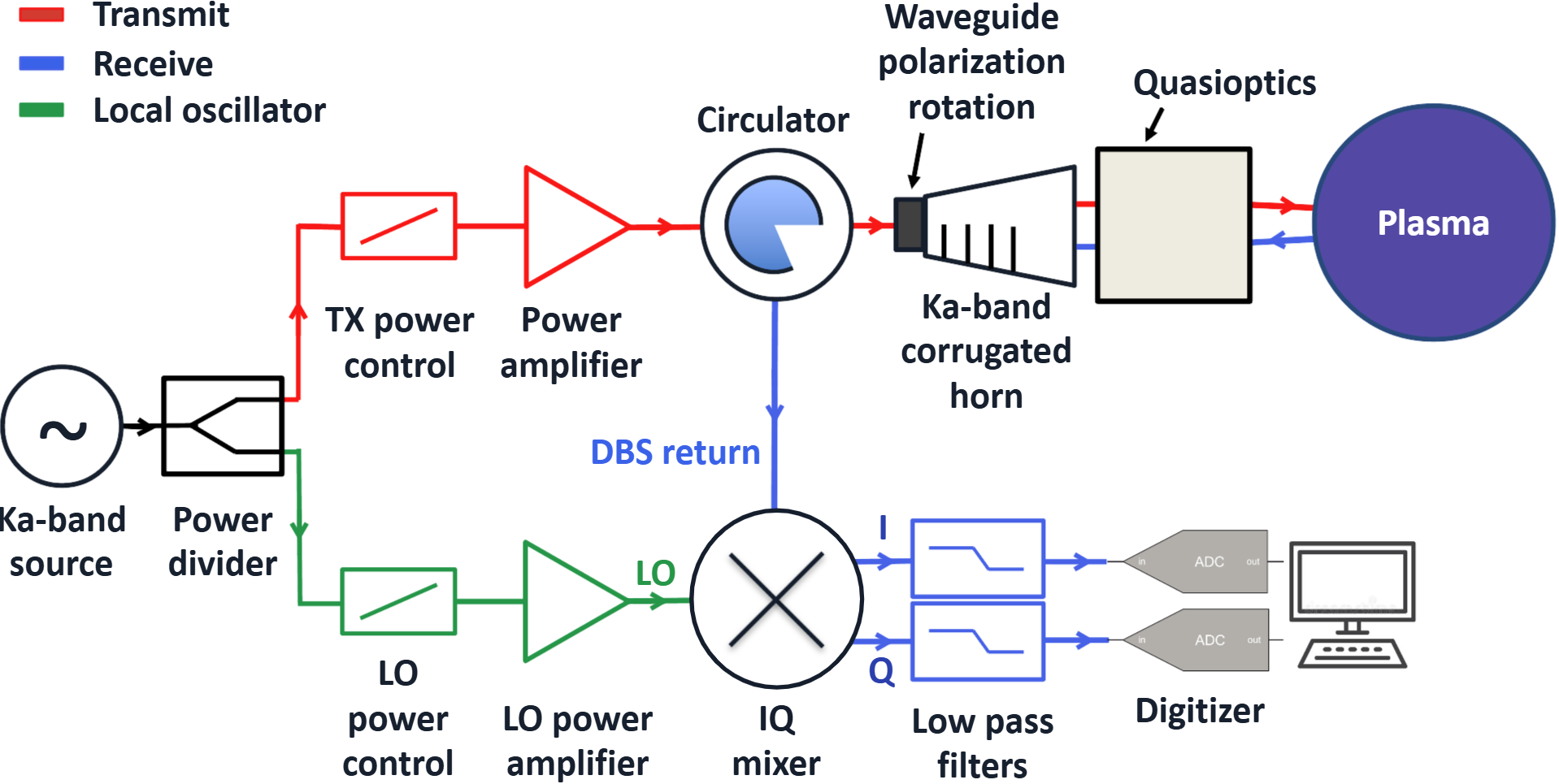}
    \caption{Functional block diagram of the proposed coherent homodyne Ka-band DBS system.}
    \label{fig:block_diagram}
\end{figure}

The proposed Ka-band DBS system uses a coherent homodyne architecture. A Ka-band microwave source is divided into a transmit branch and a local-oscillator reference branch, see Figure \ref{fig:block_diagram}. In the transmit branch, the probe power is adjusted, amplified, and routed through a circulator to a corrugated horn. The horn launches the probe beam through the quasioptical system described above, and the same optical path collects the backscattered DBS signal.

In the monostatic configuration, the circulator separates the outgoing probe wave from the returning scattered signal and routes the received signal to the mixer input. The received signal is mixed with the coherent reference signal in an IQ mixer, producing in-phase and quadrature baseband signals. These outputs are low-pass filtered and digitised for Doppler spectral analysis. This architecture preserves the phase information needed to determine the sign and magnitude of the Doppler shift while keeping the microwave hardware comparatively simple.

The principal microwave and quasioptical components have been procured, and laboratory assembly and testing are underway. These tests will characterise the source power, beam profile, polarisation purity, and launch-angle calibration before installation on Pegasus-III.


%
%
\section{Conclusions}
In this paper, we presented the preliminary design of a two-dimensionally steerable DBS diagnostic for Pegasus-III. The system is designed to operate both as a conventional DBS diagnostic, measuring turbulent density fluctuations and perpendicular flows, and as a test-bed for pitch angle DBS in a spherical tokamak.

Using representative Pegasus-III equilibria and the \textit{Scotty} beam-tracing code, we evaluated the accessible cutoff locations, measured fluctuation wavenumbers, and mismatch attenuation for a range of launch frequencies, launch angles, and probe beam polarisations. We found that a Ka-band DBS system with tuneable frequency channels and two-dimensional steering can access useful scattering locations and ion-scale fluctuation wavenumbers in Pegasus-III plasmas. We also showed that toroidal steering is required both to minimise mismatch attenuation and to resolve the toroidal response needed for pitch angle DBS.

Finally, we presented a preliminary quasioptical and hardware design capable of satisfying these requirements within the available Pegasus-III port geometry. The proposed system uses a tuneable Ka-band source, a corrugated horn, a rotatable polariser, and a two-dimensionally steerable mirror, providing a comparatively simple DBS platform while retaining flexibility in launch geometry and polarisation. The principal components have been procured, and laboratory testing is underway to characterise the microwave chain, beam properties, and launch-angle calibration.

This diagnostic will support studies of turbulence, flows, magnetic pitch angle, and O-X-B mode conversion in Pegasus-III non-inductive plasmas. More broadly, this work illustrates how relatively simple DBS hardware can be extended toward pitch angle measurements through two-dimensional steering and toroidal-response analysis. Together with existing microwave diagnostics such as SAMI, the proposed DBS system also provides a flexible platform for developing microwave diagnostic techniques relevant to spherical tokamaks, burning plasmas, and future fusion power plants.

\begin{acknowledgments}
This work was partially funded by A*STAR: Strategic Tokamak Research for Industrial Deployment, and Energy (STRIDE) grant [H26-MSE152] and the FEAT-SRTT. The authors thank M. W. Bongard for contributing the KFIT equilibrium reconstruction, T. N. Tierney and J. A. Reusch for contributing the Thomson scattering measurements, and the Pegasus-III team for operating and maintaining the facility. The Pegasus-III Experiment is based on work supported by the U.S. Department of Energy, Office of Science, Office of Fusion Energy Sciences, under Award No. DE-SC0019008. Any opinions, findings, and conclusions or recommendations expressed in this paper are those of the authors and do not necessarily reflect the views of the U.S. Department of Energy. The authors have no conflicts to disclose.

\textbf{Data availability statement}. The data that supports the findings of this study are available from the corresponding author upon reasonable request.

\textbf{AI declaration}. Generative AI tools were used to assist with language editing, paragraph restructuring, and improving the clarity and flow of the manuscript. They were also used to suggest alternative phrasings for the introduction, figure captions, section transitions, and conclusion. The scientific content, diagnostic design, beam-tracing calculations, quasioptical modelling, figures, interpretation of results, and final technical judgements were produced and verified, without AI, by the authors. AI tools were not used to generate original data, perform simulations, make figures, write the initial draft of the manuscript, nor draw scientific conclusions.
\end{acknowledgments}

\bibliography{aipsamp}

\end{document}